\documentclass[letterpaper]{article}

\usepackage{mathptmx}   
\usepackage{amsfonts, amssymb, yhmath}
\usepackage{epsfig}
\usepackage{theorem}
\usepackage{algorithmic}
\usepackage{algorithm}
\usepackage{textcomp}
\usepackage{multirow}

\newtheorem{theorem}{Theorem}

\newtheorem{corollary}[theorem]{Corollary}

\newtheorem{lemma}[theorem]{Lemma}
\newtheorem{problem}[theorem]{Problem}

\newcommand{\qed}{\hfill $\Box$\\}

\def\pmt{\textrm{PMT}}
\def\pmg{\textrm{PMG}}
\def\ppg{\textrm{PPG}}
\def\ppt{\textrm{PPT}}

\def\tec{\textrm{MTEC}}

\begin{document}

\date{}
\title{A Linear Time Algorithm for the Feasibility of Pebble Motion on Graphs\thanks{Jingjin Yu is with the Department of Electrical and Computer Engineering, University of Illinois at Urbana-Champaign, Urbana, IL 61801 USA. E-mail: jyu18@uiuc.edu. This research was supported in part by NSF grant 0904501 (IIS Robotics), NSF grant 1035345 (Cyberphysical Systems), and MURI/ONR grant N00014-09-1-1052.}
\author{Jingjin Yu}
}
\maketitle

\begin{abstract} Given a connected, undirected, simple graph $G = (V, E)$ and $p \le |V|$ pebbles labeled $1, \ldots, p$, a configuration of these $p$ pebbles is an injective map assigning the pebbles to vertices of $G$. Let $S$ and $D$ be two such configurations. From a configuration, pebbles can move on $G$ as follows: In each step, at most one pebble may move from the vertex it currently occupies to an adjacent unoccupied vertex, yielding a new configuration. A natural question in this setting is the following: Is configuration $D$ reachable from $S$ and if so, how? We show that the feasibility of this problem can be decided in time $O(|V| + |E|)$. 
\end{abstract}

\section{Introduction}

\begin{figure}[htp]
\begin{center}
  \begin{tabular}{ccc}
    \includegraphics[width=0.8in]{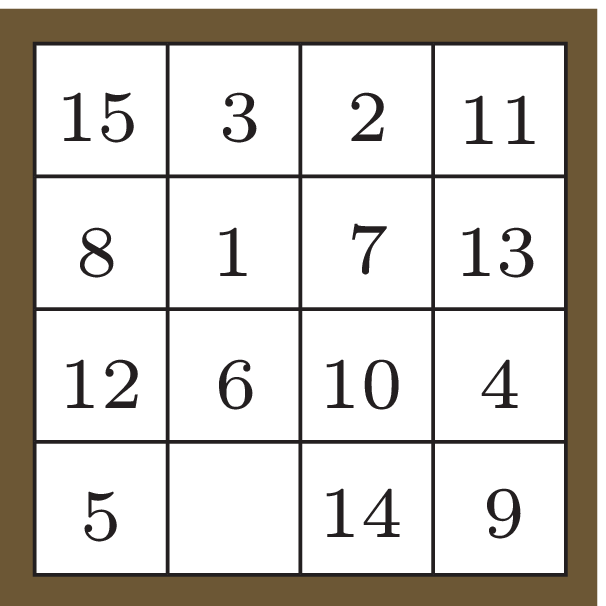} & \hspace{10mm} &
    \includegraphics[width=0.8in]{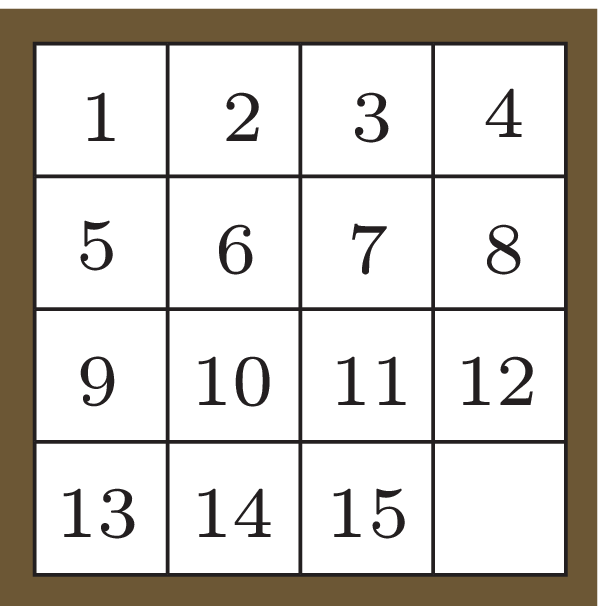}  \\
    (a) & & (b)\\
  \end{tabular}
\end{center}
\caption{\label{fig:15-puzzle} Two 15-puzzle instances. a) An unsolved instance. In the next step, one of the pieces labeled 5, 6, 14 may move to the vacant cell, leaving behind it another vacant cell for the next move. b) The solved instance.}
\end{figure}

In Sam Loyd's 15-puzzle \cite{Loy59}, a player is asked to arrange square game pieces labeled 1-15, scrambled on a $4 \times 4$ grid, to a shuffled row major ordering, using one empty swap cell: In each step, one of the labeled pieces neighboring the empty cell may be moved to the empty cell (see, e.g., Fig. \ref{fig:15-puzzle}). As early as 1879, Story \cite{Sto1879} observed that the feasibility of a 15-puzzle instance is solely decided by the parity of the starting configuration (with respect to the fixed goal configuration).

Generalizing the 15-puzzle to having $n-1$ labeled pebbles on an arbitrary non-separable graph $G$ with $n$ vertices, Wilson \cite{Wil74} formalized the observation of Story by showing that if $G$ is bipartite, the reachable configurations from a given start configuration form an alternating group on $n-1$ letters, partitioning all possible configurations into two equivalence classes. If $G$ is non bipartite (except for a special $\theta_0$ graph with seven vertices), then the reachable configurations form the symmetric group on $n - 1$ letters, implying that the problem is always feasible. The result by Wilson also yields an algorithm for solving a given instance. However, the number of moves involved may be exponential. Taking a further step, Kornhauser, Miller, and Spirakis \cite{KorMilSpi84} studied the {\em pebble motion problem} on an arbitrary connected $n$-vertex graph with up to $n - 1$ pebbles. For this problem, they gave a polynomial time algorithm that produces solutions with a $O(n^3)$ upper bound on the number of moves. 

As pointed out in \cite{KorMilSpi84}, certain instances of the pebble motion problem require $\Theta(n^3)$ moves, suggesting an $\Omega(n^3)$ lower bound on any algorithm that computes a step-by-step plan for moving the pebbles. Since not all instances of the pebble motion problem can be solved, if the feasibility test can be performed faster than $\Theta(n^3)$, unnecessary computation on infeasible instances can be avoided. Auletta et al. \cite{AulMonParPer99} showed that for trees, deciding whether an given instance of the pebble motion problem is feasible can be done in linear time. Recently, Goraly and Hassin \cite{GorHas10} extended the result to graphs. We independently reach the same conclusion via a direct reduction to the tree approach proposed in \cite{AulMonParPer99}. The tree that we obtain shrinks the graph significantly wheres the method in \cite{GorHas10} adds more vertices to the graph. 

The pebble motion problem on graphs finds applications in multi-robot path planning, deflection routing in data networks, and memory management in distributed systems. Fast feasibility test for this problem can help eliminate infeasible instances, thus avoiding unnecessary computation on parts of these instances.

\subsection{Problem Statement and Main Result}
Let $G = (V, E)$ be a connected, undirected, simple graph with $|V| = n$. The presented results readily generalize to graphs with multiple connected components (via taking the direct product of the groups from the components. Generalizations to directed graphs and graphs with multiple edges are also straightforward. Let there be a set $p \le n $ pebbles, numbered $1, \ldots, p$, residing on distinct vertices of $G$. A {\em configuration} of these pebbles is an injective map $S: \{1, \ldots, p\} \to V$. A configuration can also be viewed as a sequence of vertices, $S = \langle s_1, \ldots, s_p \rangle$. We use $V(S)$ to denote the range of $S$. A {\em move} is a pair of configurations, $\langle S, S' \rangle$, such that $S, S'$ differ at exactly one pebble $i$ and $(s_i, s_i') \in E$. That is, in a move, a single pebble may migrate from its current vertex to an empty neighboring vertex. 

When two configurations $S$ and $S'$ are parts of a move, they are {\em connected}. Two configurations $S$ and $S'$ are also connected (and therefore {\em reachable} from each other) if there exists a sequence of configurations $\langle S=S_0, \ldots, S_t = S'\rangle$ such that every pair of consecutive configurations $S_i, S_{i+1}$ in the sequence are connected. The problem of {\em pebble motion on graphs} or $\pmg$ is defined as follows. 

\begin{problem}[$\pmg$]\label{pm} Given an instance $I = (G, S, D)$ in which $G$ is a connected graph, and $S$ and $D$ are two pebble configurations on $G$, find a sequence of moves that connects configurations $S$ and $D$.
\end{problem}

It is clear that a given $\pmg$ instance may not have a solution. If it does, then it is {\em feasible}. When $G$ is a tree, $\pmg$ is also referred to as {\em pebble motion on trees} ($\pmt$). In this case, an instance is usually written as $I = (T, S, D)$. Auletta et al. \cite{AulMonParPer99} showed that the feasibility test of a $\pmt$ instance can be performed in $O(n)$ operations. We generalize this linear time result to graphs. The main result of the paper is presented in the following theorem.  

\begin{theorem}\label{t:main} The feasibility test of $\pmg$ can be performed in $O(|V| + |E|)$ time.
\end{theorem} 

The key idea behind our linear time feasibility test is reducing a $\pmg$ instance to a $\pmt$-like instance, allowing many ideas from \cite{AulMonParPer99} to be adapted for proving Theorem \ref{t:main}. This leads to some intermediate results that look similar to those from \cite{AulMonParPer99} but require significantly different proofs. To make this paper self contained, complete proofs are generally provided for these intermediate results. 

\section{Reducing Pebble Motion on Graphs to Pebble Permutation on Graphs}

If $V(S) = V(D)$ in a $\pmg$ instance $(G, S, D)$, $D$ can be viewed as a {\em permutation} $\Pi$ of $S$, defined as $d_i = s_{\Pi(i)}$ for all $i$ (i.e., $\Pi$ permutes the pebbles). We call such a problem {\em pebble permutation on graphs}, or $\ppg$. The main goal of this section is to show that any $\pmg$ instance can be reduced to an equivalent $\ppg$ instance such that the $|V(S)|$ pebbles can occupy any set of $|V(S)|$ vertices on $G$.

If the underlying graph is a tree in a $\ppg$ instance, the problem becomes {\em pebble permutation on trees}, or $\ppt$. Reducing a $\pmg$ instance to an equivalent $\ppg$ can be done using Theorem 3 from \cite{AulMonParPer99}, a restatement of which is given below. We give a shorter constructive proof of this result. 

\begin{theorem}\label{t:pmt-ppt} Let $(T, S, D)$ be a $\pmt$ instance. In $O(n)$ steps, an instance $(T, S', \Pi)$ of \ppt\, can be computed such that $S, S'$ are connected and for all $i$, $d_i = s_{\Pi(i)}'$ for a fixed permutation $\Pi$.
\end{theorem}
{\sc Proof.} We produce a mapping between $S$ and a new configuration $S'$ so that the requirements are satisfied. To start, all leaf vertices of $T$ are put into a queue $Q$ and {\em processed} in the order they are added. After a vertex $v$ from $Q$ is processed, its neighbors, $N(v)$, are examined. If a neighbor $u \in N(v)$ has not been added to $Q$, $u$ is added to $Q$ if $N(u)$ has at most one member which has not already been added to $Q$. It is straightforward to check that adding vertices to $Q$ this way guarantees that $Q$ will not be empty until all vertices of $T$ are processed. 

The processed vertices form a forest, $F$, of which the trees eventually combine to yield $T$. In this proof, $v$ is always assumed to be the current vertex from $Q$ that is being processed and is adjacent to the tree $T_i \in F$ (this does not prevent $v$ from being adjacent to other trees from $F$). As $v$ is being processed, $T_i$ will be examined. Depending on how $|V(S) \cap V(T_i)|$ and $|V(D) \cap V(T_i)|$ compare, there are three possibilities. 

First, if $|V(S) \cap V(T_i)| = |V(D) \cap V(T_i)|$, then nothing additional is needed to be done for $T_i$. 

Next, if $|V(S) \cap V(T_i)| > |V(D) \cap V(T_i)|$, then some pebbles will need to be moved out of $T_i$ through its root so that the numbers of pebbles on $T_i$ from $S$ and $D$ are the same. For such a tree $T_i$, a {\em surplus} queue, $Q_i^+$, of pebbles will be maintained, so that pebbles at the front of the queue are readily moved out of the root of $T_i$. To maintain $Q_i^+$, the operations for removing and adding a pebble, as well as merging of two queues need to specified (one more operation involving a surplus queue will be introduced in the next paragraph). Removing a pebble from $Q_i^+$ is needed when $v \in V(D)$ and $v \notin V(S)$; a pebble from $Q_i^+$ needs to move to $v$. For this, simply grab the pebble at the end of $Q_i^+$, since that pebble can be the last pebble to leave the current $T_i$. To add a pebble (needed when $v \in V(S), v \notin V(D)$), insert the pebble in the front of $Q_i^+$. Note that it is possible that $v \in (V(S) \cap V(D))$; in this case the removal is followed by the insertion. If $v$ will be the new root of two trees $T_i, T_j$ and both trees have surplus queues ($Q_i^+, Q_j^+$, respectively), before processing $v$, merge $Q_i^+, Q_j^+$ by attaching $Q_j^+$ at the end of $Q_i^+$. This works since pebbles from $Q_i^+, Q_j^+$ must all move through $v$; all pebbles in $Q_i^+$ can move through $v_i$ first. Same applies if there are three or more trees meeting at $v$. 

Finally, if $|V(S) \cap V(T_i)| < |V(D) \cap V(T_i)|$, pebbles will need to be moved into $T_i$ through its root. A {\em deficit} queue $Q_i^-$ containing vertices of $V(D)$ is maintained in this case. Assuming $Q_i^-$ is arranged such that the front vertex is close to the root of $T_i$. The operations for removing, adding, and merging deficit queues mirror those for surplus queues. An extra queue operation needed here is when a deficit queue $Q_i^-$ needs to be merged with a surplus queue $Q_j^+$; this can be done simply by filling $Q_i^-$ from the back with pebbles of $Q_j^+$ starting from the front. 
~\qed

The constructive proof can be easily adapted to yield paths for actually moving the pebbles (note that doing this requires more than linear time). Theorem \ref{t:pmt-ppt} allows us to reduce a $\pmg$ to an equally feasible $\ppg$. 

\begin{corollary}\label{c:reduce} An instance $I = (G, S, D)$ of $\pmg$ can be reduced to a $\ppg$ instance, $I' = (G, S', \Pi)$, in linear time.  
\end{corollary}
{\sc Proof.} Given a $\pmg$ instance $I = (G, S, D)$, compute in linear time a spanning tree $T_G$ of $G$. From the $\pmt$ instance $(T_G, S, D)$, compute a equally feasible $\ppt$ instance $(T_G, S', \Pi)$ (via Theorem \ref{t:pmt-ppt}) in which $d_i = s_{\Pi(i)}'$ for all $i$. $I' = (G, S', \Pi)$ is the desired $\ppg$ instance. To see that $I$ and $I'$ are equally feasible, note that the sequence of configurations connecting $S, S'$ in $T_G$ are still present on $G$. ~\qed

Corollary \ref{c:reduce} yields another useful corollary, which says that a $\ppg$ instance can be converted to an equivalent one such that the pebbles occupy an arbitrary set of vertices. 

\begin{corollary}\label{c:reduce2} Let $I = (G, S, \Pi)$ be an arbitrary $\ppg$ instance and let $V_A$ be an arbitrary set of $|V(S)|$ vertices of $G$. Then $I$ can be reduced to a $\ppg$ instance, $I' = (G, S', \Pi)$, in linear time, such that $V(S') = V_A$.
\end{corollary}
{\sc Proof.} Let $A$ be an arbitrary configuration of the $|V(S)|$ pebbles with $V(A) = V_A$. From the $\pmg$ instance $(G, S, A)$, a $\ppg$ instance $(G, S', \Pi')$ can be computed by Corollary \ref{c:reduce} such that the $\pmg$ instance $(G, S, S')$ is feasible.  Let $D$ be the goal configuration of $I$ (i.e., $d_i = s_{\Pi(i)}$). Applying the same moves (that take $S$ to $S'$) to $D$ yields $D'$ satisfying $d_i' = s_{\Pi(i)}'$. Therefore, $S'$ occupy the same set of vertices as $A$ and $(G, S, \Pi)$ is feasible if and only if $(G, S', \Pi)$ is feasible. ~\qed

\section{Partitioning of $\ppg$ Instances}\label{sec:partition}

We now partition a $\ppg$ instance $I = (G, S, \Pi)$ based on the graph $G$ as some cases require relatively simple but special treatment. A {\em maximal $2$-edge-connected component} ($\tec$ for short) of $G$ is a $2$-edge-connected component of $G$ that is not contained in any other $2$-edge-connected component of $G$. We use $n_{M}$ to denote the number of vertices of all $\tec$s of $G$. The $\tec$s of a graph can be found in $O(|V| + |E|)$ time \cite{Tar72}. The main goal of this section is to solve all $\ppg$ instances other than these with $n_M - 2 \le p \le n - 2$. The case of $p = n$ is trivial. For the discussion in this section, unless otherwise stated, let $(G, S, D)$ be the $\pmg$ instance identical to $I$ (that is, $d_i = s_{\Pi(i)}$). 

The first special case is when $G$ is the $\theta_0$ graph with seven vertices, which is formed by connecting an extra vertex to two vertices of distance $3$ on a hexagon \cite{KorMilSpi84}. Any $\ppg$ instance in which $G$ is the $\theta_0$ graph can be solved in constant time since there are only a finite number of possible configurations. 

The second special case is when $G$ is a cycle. In this case, $S$ and $D$, as sequences of vertices, induce natural cyclic orderings of the pebbles. This implies that $I$ is feasible if and only if $s_i = d_{(i + k) \textrm{ mod } k}$ for some fixed natural number $k$. The associated computation requires linear time. 

The third special case is when $G$ is a $2$-connected (i.e., non separable) graph that is not a cycle or the $\theta_0$ graph. As pointed out in \cite{Wil74}, if $G$ is bipartite and $p = n - 1$, all configurations of pebbles fall into two equivalence classes. In each equivalence class, all configurations are connected and two configurations from different configurations classes are not connected. Deciding whether two configurations are connected can be performed in linear time by computing the {\em parity} of the two configurations (i.e., treating the configurations as permutations). Continuing on this special case, if $G$ is bipartite and $p \le n - 2$, or if $G$ is not bipartite and $p \le n -1$, then all configurations are connected. We summarize the cases mentioned so far in the following lemma.

\begin{lemma}\label{l:special} Let $I = (G, S, \Pi)$ be a $\ppg$ instance in which $G$ is a $2$-connected graph. The feasibility test of $I$ can be performed in linear time. Moreover, $I$ is always feasible if: 
\begin{enumerate}
\item $G$ is not a cycle and $p \le n - 2$, or
\item $G$ is not a cycle, not the $\theta_0$ graph, and $p \le n - 1$. 
\end{enumerate}\end{lemma}

If a $\ppg$ instance is always feasible, it means that any pair of pebbles can switch locations without affecting other pebbles. In general, when two pebbles can exchange locations without net effects on the locations of other pebbles, they are {\em equivalent} with respect to the specific configuration they are associated with. More formally, two pebbles $i, j$ are equivalent with respect to a configuration $S$ if $S$ is connected to a configuration $S'$ in which $s_i = s_j', s_j = s_i'$, and $s_k = s_k'$ for all $k \ne i, j$. A set of pebbles are equivalent if every pair of pebbles from the set are equivalent. It is clear that this type of equivalence is reflexive and transitive. Note that our definition of pebble equivalence equals the definition of vertex equivalence used in \cite{AulMonParPer99}. 

\begin{corollary}\label{c:feasible-0} Let $I = (G, S, \Pi)$ be an instance of $\ppg$ in which $G$ is a single $2$-connected component plus a single degree one vertex attached to that component. If $p \le n - 2$, then $I$ is feasible. 
\end{corollary} 
{\sc Proof.} Let the component be $H$ and the degree one vertex be $v$. We may assume that in $S$, $v$ is occupied by a pebble $i$. If $H$ is a cycle, with two empty vertices on the cycle $H$, it is straightforward to check that $i$ can be exchanged with any other pebble on $H$. This implies that all pebbles are equivalent with respect to $S$. Thus, $I$ is feasible for an arbitrary $\Pi$. 

If $H$ is not a cycle, with pebble $i$ on $v$, all pebbles on $H$ are equivalent with respect to $S$ by Lemma \ref{l:special}. Fixing any cycle $C$ on $H$ adjacent to $v$, moving the empty vertices to $C$ shows that pebble $i$ is equivalent to at least one pebble on $C$. By transitivity, all pebbles are again equivalent with respect to $S$. ~\qed

The fourth special case is when $G$ is $2$-edge-connected and separable (i.e., not $2$-connected), for which statements similar to that from Lemma \ref{l:special} can be made. We include all four cases in the following theorem. 

\begin{theorem}\label{t:special}Let $I = (G, S, \Pi)$ be a $\ppg$ instance in which $G$ is a $2$-edge-connected graph. The feasibility test of $I$ can be performed in linear time. Moreover, $I$ is always feasible if $G$ is not a cycle and $p \le n - 2$.
\end{theorem}
\begin{figure}[htp]
\begin{center}
    \includegraphics[width=2in]{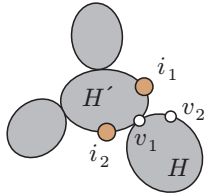} 
\end{center}
\caption{\label{fig:2c} A $2$-edge-connected graph can be viewed as a tree of its maximal $2$-connected components.}
\end{figure}
{\sc Proof.} Since the case of $G$ being non separable is covered by Lemma \ref{l:special}, assume that $G$ is separable. $G$ is then two or more $2$-connected graphs joined at articulation vertices, forming a tree like structure (see, e.g., Fig. \ref{fig:2c}). If $p \le n - 2$, without loss of generality, assume (given configuration $S$) that there are two empty vertices $v_1, v_2$ on a ``leaf'' $2$-connected component $H$ of $G$. By Corollary \ref{c:feasible-0}, the current pebbles on $H$ in $S$ and $i_1, i_2$ are all equivalent. Restricting our attention to $H'$ and $v_2$, all pebbles on $H$' are equivalent with respect to $S$ by Corollary \ref{c:feasible-0}. By transitivity of pebble equivalence, all pebbles on $H, H'$ in $S$ are equivalent. Inductively, all pebbles in $S$ are equivalent.

If $p = n - 1$, imagine in Fig. \ref{fig:2c} that (given configuration $S$) all vertices other than $v_1$ are occupied. Since no pebbles can cross the border between $H, H'$, if $I|_H$ ($I$ restricted to $H$ in the natural way) is not feasible, then $I$ cannot be feasible. Same applies to $I|_{H'}$. By shifting the empty vertex to other $2$-connected components, additional restricted instances can be obtained. $I$ is feasible if and only if all such restrictions are feasible. Since these instances can be computed independently, observe that the overall time needed is $O(|V| + |E|)$.~\qed

With Theorem \ref{t:special}, we can state a more useful version of Corollary \ref{c:feasible-0}.

\begin{corollary}\label{c:feasible} Let $I = (G, S, \Pi)$ be an instance of $\ppg$ in which $G$ is a single $\tec$ with a single degree one vertex attached to the $\tec$. If $p \le n - 2$, then $I$ is feasible. 
\end{corollary} 

Next, we look at the case of $p = n - 1$ for general graphs. The proof strategy is similar to that used for proving the $p = n - 1$ case of Theorem \ref{t:special}.

\begin{theorem}The feasibility of a $\ppg$ instance in which $p = n - 1$ can be decided in $O(|V| + |E|)$ time. \end{theorem}
{\sc Proof.} Let $I = (G, S, \Pi)$ be an instance of $\ppg$ in which $p = n-1$. The claim holds when $G$ is a tree or a $2$-edge-connected graph; assume $G$ is not such a graph. Let $H$ be an $\tec$ of $G$. Without loss of generality, we may assume that in configuration $S$, the only unoccupied vertex is within $H$, leaving $|V(H)| - 1$ pebbles on $H$. By Theorem \ref{t:special}, the feasibility of $I|_H$ can be decided in linear time.  

If $I|_H$ is not feasible, then $I$ itself cannot be feasible. If $I|_H$ is feasible, then other $\tec$s of $G$ are examined next. For this, configurations $S$ is updated to $S'$such that another $\tec$ (if any) of $G$ will now have one unoccupied vertex (note that pebbles do not need to be actually moved). Perform the same movements on $D$ gives $D'$ with the same unoccupied vertex (recall that $D$ is the goal configuration). This yields an equivalent $\ppg$ instance $I' = (G, S', \Pi)$. Let the $\tec$ with the unoccupied vertex be $H'$, feasibility of $I'|_{H'}$ can also be checked in linear time. If any $I'|_{H'}$ is infeasible, then $I$ is infeasible. Checking the feasibility of all such restricted instances take total time $O(|V| + |E|)$. 

If $I$ remains feasible after above checks, rest of the pebbles (those that have not appeared in $I|_H$ or an $I'|_{H'}$) must be examined. For each of these pebbles, say pebble $i$, if $s_i \ne d_i$, then $I$ is not feasible. Otherwise, $I$ is feasible. ~\qed

The last special case is when $p \le n_{M} - 3$. 

\begin{theorem}\label{t:nmminus3} A $\ppg$ instance is feasible when $p \le n_M - 3$ and $G$ is not a cycle. \end{theorem}
{\sc Proof.} Let $I = (G, S, \Pi)$ be an arbitrary instance of $\ppg$ in which $p \le n_M -3$. This excludes $G$ from being a tree. If $G$ is a $2$-edge-connected graph and not a cycle, then the claim trivially holds by Theorem \ref{t:special}. Same is true if $G$ contains only one $\tec$ by Corollary \ref{c:feasible}. For the rest of the proof, assume that $G$ contains two or more $\tec$s. We only work with configuration $S$ and prove the case $p = n_M - 3$. Other cases then trivially follow. 

\begin{figure}[htp]
\begin{center}
    \includegraphics[width=1.6in]{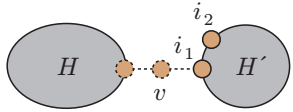} 
\end{center}
\caption{\label{fig:nmminus3} Moving two pebbles between to adjacent $\tec$s $H$ and $H'$. The two small dotted circles are the new locations of pebbles $i_1$ (left) and $i_2$ (right).}
\end{figure}

By Corollary \ref{c:reduce2}, assume without loss of generality that all $p$ pebbles are on $\tec$s of $G$ and one $\tec$, say $H$, has at three unoccupied vertices. This means that $\tec$s other than $H$ are fully occupied. Let $H'$ be another $\tec$ such that there are no other $\tec$s between $H$ and $H'$. Let $i_i, i_2$ be two pebbles on $H'$ that occupy vertices closest to $H$. Since there are no pebbles between $H$ and $H'$, pebbles $i_1, i_2$ can be moved such that $i_1$ is on $H$ and $i_2$ is on a vertex adjacent to $H$ (see e.g., Fig. \ref{fig:nmminus3}). Let the vertex to which $i_2$ is moved be $v$; note that $v$ may be on $H'$. Let the new configuration be $S'$. By Corollary \ref{c:feasible}, the pebbles current on $H$ plus $i_2$ are all equivalent with respect to $S'$ and therefore, $S$. 

On the other hand, from the configuration $S$, $i_1$ can be moved away from $H'$ and $i_2$ can then be moved to a vertex adjacent to $H'$. This leaves two empty vertices on $H'$; let this configuration be $S''$. Following the same argument, all pebbles on $H'$ plus $i_2$ are equivalent with respect to $S''$ and therefore, $S$. By transitivity, all pebbles on $H, H'$ in configuration $S$ are equivalent. Inductively, this shows that all pebbles are equivalent with respect to $S$. Thus $I$ is feasible. 
~\qed

\section{Reducing Pebble Permutations to Pebble Exchanges}

In this section, let $I = (G, S, \Pi)$ be a $\ppg$ instance in which $G$ contains an $\tec$, $G$ is not a single $\tec$, and $p \le n - 2$. We show that such an instance is feasible if and only if for all $i$, pebbles $i$ and $\Pi(i)$ are equivalent with respect to $S$. That is, pebbles $i$ and $\Pi(i)$ can be exchanged without net effects on other pebbles. 

\begin{lemma}\label{l:ij} Let $S$ be an arbitrary configuration and let $i, j$ be two pebbles. If both $i$ and $j$ can reach two distinct vertices (not necessarily the same two vertices for $i, j$) of an $\tec$ $H$, then $i$ and $j$ are equivalent with respect to $S$.
\end{lemma}
{\sc Proof.} Assume without loss of generality that pebble $i$ first gets to a configuration $S'$ such that it is on $H$ and can move to a nearby empty vertex on $H$. Assume that pebble $j$ can do the same in a later configuration $S''$ (if both pebbles $i, j$ are already on $H$ with an empty vertex on $H$, $i, j$ are equivalent by Corollary \ref{c:feasible}). 

Starting from $S'$, if there is only one empty vertex on $H$, move pebbles outside $H$ such that an vertex adjacent to $H$ is empty (since $G$ is not a single $\tec$). This setup satisfies the conditions of Corollary \ref{c:feasible}. Therefore, all pebbles on $H$, including $i$, are equivalent with respect to the current configuration and also $S'$. Since an $\tec$ contains at least $3$ vertices, $i$ is at least equivalent to one other pebble (note that this pebble may not be on $H$ in configuration $S'$). 

From configuration $S'$, to move $j$ to $H$, some pebbles may need to be moved to $H$ and some pebbles, including $i$, may get moved out of $H$. We now augment the configurations between $S'$ and $S''$ so that pebble $i$ never leaves $H$. First note that if $i$ is the only pebble on $H$ and is supposed to leave $H$, then it must be for some other pebble to move into $H$. It is straightforward to verify that the next pebble entering $H$ must also be equivalent to $i$ and can take on pebble $i$'s role. If $i$ is not the only pebble on $H$ and is supposed to leave, again we can let some pebble equivalent to $i$ leave $H$. With this augmentation, when pebble $j$ eventually gets to $H$ and can move to another vertex of $H$, $i$ and $j$ are clearly equivalent with respect to the current configuration and therefore, $S$ (exchange $i, j$ and reverse all previous moves).
~\qed

For a given $G$, if a vertex $x$ is an articulation vertex, removing $x$ splits $G$ into two or more components. Denote these components as $C(x)$ ($G$ is assumed). Denote the component from $C(x)$ containing $y$ as $C(x, y)$ and the rest $\overline{C(x, y)}$. We now give a generalization of Lemma 6 from \cite{AulMonParPer99}. 

\begin{lemma}\label{l:uvw} Let $S$ be an arbitrary configuration and suppose that $i, j$ are two pebbles occupying vertices $u, v$, respectively. Let $w$ be a vertex such that two shortest paths between $u, v$ and $v, w$ share at least one common edge. Suppose that there are moves that take $S$ to a  configuration in which $i$ is at $v$ and $j$ is at $w$. Then $i, j$ are equivalent with respect to $S$. 
\end{lemma}
\noindent {\sc Proof.} Lemma 6 from \cite{AulMonParPer99} shows that Lemma \ref{l:uvw} holds when $G$ is a tree. Our proof seeks to reduce our case to the tree case. Note that when $G$ is a tree, the shortest paths between a pair of vertices are unique, which is not the case for a general graph. The statement of Lemma \ref{l:uvw} only requires that a shortest path between some $u, v$ and a shortest path between $v, w$ share an edge. Also, pebbles other than $i, j$ can be treated as indistinguishable pebbles; their locations before and after the moves do not matter.

Let the sequence of moves that takes $i, j$ to $v, w$, respectively, be $X$. For convenience, let $u \leadsto v$ and $v \leadsto w$ denote two shortest paths that share at least one edge. These paths may not be unique on $G$. We require that $u\leadsto v$ and $v\leadsto w$ have only one intersection (which may contain multiple edges and vertices), denoted $y \leadsto v$. $u\leadsto y$ and $y \leadsto w$ denote the parts of $u\leadsto v$ and $v \leadsto w$, respectively, after removing $y \leadsto v$. Note that fixing a $u\leadsto v$, if some vertex $z$ on $u\leadsto v$ is on an $\tec$ then any path that $i$ takes to reach $v$ from $u$ must also reach the same $\tec$ even if $i$ does not pass $z$. 

If $u, v$ belong to the same $\tec$ $H$, any $u \leadsto v$, as a shortest path, must fall entirely in $H$. This suggests that $i$ and $j$ can both reach two vertices of $H$. $i, j$ are equivalent by Lemma \ref{l:ij}. Assume for the rest of the proof that $u, v$ do not belong to the same $\tec$.

If any edge $e$ in $y \leadsto v$ appear in an $\tec$ $H$, then $i, j$ must visit at least two vertices of $H$ (not necessarily passing edge $e$) and must be equivalent. We may then assume that $y\leadsto v$ has no edges belonging to an $\tec$ of $G$. Furthermore, we may assume that no vertices on $y \leadsto v$ (including $y$ but not $v$) belong to any $\tec$ $H$; otherwise $i, j$ must be equivalent. To see this, assume without loss of generality that $y$ is on some $\tec$ $H$. At some point, $i$ must pass through $H$. If $i$ travels through an edge of $H$, we return to the earlier case; suppose not. This forces $i$ to pass $H$ through $y$. When $i$ has just arrived at $y$, there is an empty vertex on $u \leadsto y$ and there must also be an empty vertex in $\overline{C(y, u)}$. These two empty vertices allow $i$ to reach at least one more vertex of $H$ other than $y$. Same applies to $j$, making $i, j$ equivalent with respect to $S$ by Lemma \ref{l:ij}. 

\begin{figure}[htp]
\begin{center}
    \includegraphics[width=2in]{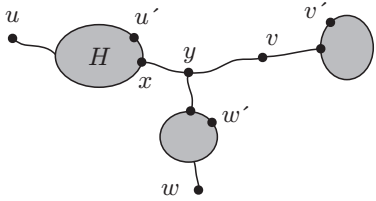} 
\end{center}
\caption{\label{fig:pebble-equ} If $i, j$ (on $u, v$, respectively) can reach $v, w$, then a pebble equivalent to $i$ can reach $v$ without using features of $\tec$ as $j$ moves to $w$ .}
\end{figure}

We now look at $u \leadsto y$. If any vertex inside $u \leadsto y$ (i.e., excluding $u, y$) belong to some $\tec$, let the $H$ be such an $\tec$ that is closest to $y$ ($H$ may not be unique). Since $i$ can reach $H$, it can reach two vertices of $H$ (by the argument in the previous paragraph). Pebble $i$ can then take take any position on $H$. Let $x$ be the vertex of $H$ that is closest to $y$ (this vertex is unique), we may assume that $i$ is initially located on a vertex $u'$ on $H$ adjacent to $x$ (see e.g., Fig. \ref{fig:pebble-equ}). Using the argument in the proof of Lemma \ref{l:ij}, $X$ can be modified so that $i$ never needs to move to vertices other than $u'$ in $\overline{C(x, y)}$. This means that we can effectively treat $x$ as a $T$-junction instead of a vertex of an $\tec$ for moving $i, j$ to $v, w$, respectively. If no vertices inside $u \leadsto y$ is on an $\tec$, it is possible that $X$ requires that $i$ to travel in the opposite direction of $u \leadsto y$, further away from $u$. Same argument shows that a similar $u'$ exists (e.g., $v'$ in Fig. \ref{fig:pebble-equ}). 

Following the same argument, $v', w'$ can be defined similarly so that $X$ can be modified to take $i$ from $u'$ to $v$ and $j$ from $v$ to $w'$, without $i, j$ ever reaching two vertices of any $\tec$ other than $u', v'$, and $w'$. Since no features of $\tec$s are used, we return to the tree case and $i, j$ are equivalent. ~\qed

With Lemma \ref{l:uvw}, Corollaries 1 and 2 from \cite{AulMonParPer99} can be extended to general graphs. We only need the extended version of Corollary 1, stated and proved below. 

\begin{corollary}\label{c:exchange}Let $I = (G, S, \Pi)$ be a feasible $\ppg$ instance. Then there exists $1 \le i \le p$, such that pebbles $i$ and $\Pi(i)$ are equivalent with respect to $S$. 
\end{corollary}
{\sc Proof.} Applying the proof of Corollary 1 from \cite{AulMonParPer99} to a spanning tree of $G$. ~\qed

The main result of this section is an extension of Theorem 4 in \cite{AulMonParPer99} to general graphs. We provide a shorter proof enabled by the transitivity of pebble equivalence. 

\begin{theorem}\label{t:equiv}Let $I = (G, S, \Pi)$ be a feasible $\ppg$ instance. Then for all $1 \le i \le p$, pebbles $i$ and $\Pi(i)$ are equivalent with respect to $S$. 
\end{theorem}
{\sc Proof.} After one application of Corollary \ref{c:exchange}, at least one pebble can be moved to its desired goal vertex (assuming that $\Pi$ is not the identity permutation). Repeated applications of pebble exchanges will eventually move pebble $i$ to vertex $s_{\Pi(i)}$. By transitivity of pebble equivalence, $i$ can be exchanged with a pebble occupying $s_{\Pi(i)}$ at some point. Similarly, any pebble exchanged to occupy $s_{\Pi(i)}$ at any point must be equivalent to $\Pi(i)$. Thus, $i$ and $\Pi(i)$ are equivalent with respect to $S$. ~\qed

An implication of Theorem \ref{t:equiv} is that to test the feasibility of a $\ppg$ instance, $I = (G, S, \Pi)$, we may work with $S$ and $\Pi$ separately by first computing pebble equivalence classes with respect to $S$. Two pebbles are put into the same class if and only if they are equivalent with respect to $S$. The instance $I$ is feasible if and only if pebbles $i$ and $\Pi(i)$ belong to the same equivalence class by Theorem \ref{t:equiv}. 

\section{Linear Feasibility Test of Pebble Exchanges}

In this section, we reduce the feasibility test of pebble exchanges on general graphs to the feasibility test of pebble exchanges on trees in linear time. The linear time algorithm from \cite{AulMonParPer99} then applies. The key lemma enabling this reduction is as follows. 

\begin{lemma}\label{l:shrunk} Let $I = (G, S, \Pi)$ be a $\ppg$ instance in which $p \le n - 2$ and $G$ contains an $\tec$ $H$ with one empty vertex. Contract $H$ into a single edge $(v_1, v_2)$ such that all vertices adjacent to $H$ are now adjacent to $v_1$. Let this new graph be $G'$. All pebbles already on $H$ are treated as a single pebble (equivalence class) staying on $v_2$, leaving $v_1$ empty. Then the pebble equivalence classes on $G$ with respect to $S$ is the same as the pebble equivalence classes computed on $G'$.
\end{lemma}
{\sc Proof.} We need to show that movements of pebbles can be done with $H$ if and only if equivalent movements can also be done using the new structure. Call the single pebble that represents the $|V(H)| - 1$ pebbles the {\em composite} pebble. 

First, note that it is never necessary to have more than two empty vertices on $H$ in any planned moves of pebbles. To see this, suppose at some point more than two vertices of $H$ are to be emptied. The reason for doing this can only be to allow other pebbles to move into or through $H$. However, with two empty vertices on $H$, both objectives can already be achieved. To move a pebble $i$ through $H$, with two empty vertices, $i$ can enter $H$, leaving one empty vertex on $H$. This empty vertex then allows $i$ to move to any desired exit. 

Next, observe that the only reason to fill $H$ with pebbles is when a pebble need to ``pass by'' $H$ (i.e., the pebble enters and leaves $H$ without visiting other vertices of $H$). To see this, suppose a pebble $i$ enters $H$, making $H$ fully occupied. If $i$ is to move to other vertices of $H$, then some other vertices of $H$ must be emptied first, which can be done before $i$ enters $H$. 

We may now assume that the planned moves never move more than two vertices out of $H$ and $H$ is never full unless a pebble is to pass through $v_1$. This leaves three possible operations involving $H$: 1. Moving out a pebble from $H$, leaving $|V(H)| - 2$ pebbles on $H$, 2. Moving a pebble into $H$ when there are $|V(H)| - 2$ pebbles on $H$, and 3. Moving a pebble through $H$ (not a ``pass by''). 

For the first case, with $|V(H)| - 1$ pebbles on $H$, any pebble $i$ on $H$ can be moved to a desired exit. Note that this requires a vertex adjacent to $H$, say $v$, to be empty. To carry out the same operation on the edge $(v_1, v_2)$, we empty the composite pebble from $v_2$ to $v_1$ and let it represent pebble $i$ (other pebbles represented by the composite pebble stay at $v_2$ and are ``invisible''). $i$ can then be moved to $v$ as well. It is clear that the only if part is also true. The second case is the reverse of the first case. 

For the third case, suppose we want to move a pebble $i$ from a vertex $u$ adjacent to $H$ to a vertex $v$, also adjacent to $H$. For this to be doable through $H$, there must be at least two empty vertices between $H$ and $v$. Without loss of generality, assume that $H$ has one empty vertex. With two empty vertices between $H$ and $v$, we first move a pebble, say $j$, from $H$ to $v$. The two vertices allows $i$ to be exchanged with a pebble $k$ on $H$. $k$ now occupies $u$. This again leaves two empty vertices on $H$, allowing $i$ to exchange with $j$ on $v$. The $|V(H) -1|$ pebbles on $H$ can then be returned to their initial configuration by Corollary \ref{c:feasible}. The same operation is straightforward to carry out on $(v_1, v_2)$ and $u, v$: With two empty vertices, $v_2$ and $v$ can be emptied, allowing $i$ to move to $v$ directly. ~\qed

For a $\tec$ $H$ that is fully occupied, it is also converted as outlined in the above lemma. The only difference is that a pebble now needs to be put on $v_1$. This pebble cannot always be arbitrarily selected from the pebbles on $H$. Since there are at least two empty vertices somewhere on $G$, at least two pebbles can be moved out of $H$. If there are two vertices adjacent to $H$ that can be emptied (without moving pebbles on $H$), then all pebbles on $H$ are equivalent (with respect to the current configuration) by Corollary \ref{c:feasible}. To shrink $H$ in this case, an arbitrary pebble on $H$ can be selected to occupy $v_1$ and the other $|V(H)| - 1$ are combined into a composite pebble that occupies $v_2$. On the other hand, if only one vertex adjacent to $H$, say $v$, can be emptied, the pebble occupying the vertex of $H$ adjacent to $v$ may not be equivalent to the rest $|V(H)| - 1$ pebbles. The $|V(H)| - 1$ pebbles, however, are equivalent by Corollary \ref{c:feasible}. In this later case, we let this not necessarily equivalent pebble occupy $v_1$ (in the shrunk graph $G'$) and combine the rest into a composite pebble occupying $v_2$. The equivalence between $H$ and the converted edge can be formally proven using the same proof from Lemma \ref{l:shrunk} (need to add a case that moves the pebble on $v_1$ away before any other operations). 

\begin{figure}[htp]
\begin{center}
    \includegraphics[width=3.2in]{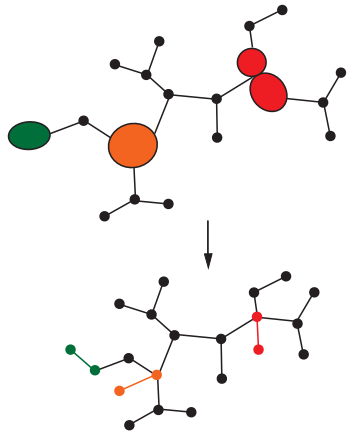} 
\end{center}
\caption{\label{fig:reduce} A graph with three $\tec$s (top) and the converted tree (bottom).}
\end{figure}

A reduction example is given in Fig. \ref{fig:reduce}.

\begin{lemma}\label{l:main} Let $I = (G, S, \Pi)$ be a $\ppg$ instance in which $n_M - 2 \le p \le n - 2$. The feasibility of $I$ can be decided in time $O(|V| + |E|)$. 
\end{lemma}
{\sc Proof.} The case of $G$ being a tree or a single $\tec$ is already covered. If $G$ contains a single $\tec$ and $p = n_M - 2$, then $I$ is feasible by Corollary \ref{c:feasible}.

For the rest of the cases, by Corollary \ref{c:reduce2}, we may assume that in configuration $S$, any $\tec$ $H$ is occupied by at least $|V(H)| -1$ pebbles. Using the reduction from Lemma \ref{l:shrunk} (and the comments that follow), all $\tec$s can be converted into single edges (with the associated pebbles combined), yielding a tree in the end. From this we can obtain a $\ppt$ instance. Grouping pebbles into equivalence classes for this $\ppt$ instance can be performed in linear time using the algorithm from \cite{AulMonParPer99}.~\qed

Combining Lemma \ref{l:main} with the results (i.e., $p \le n_M - 3$ or $p \ge n -1$) from Section \ref{sec:partition} yields the main result (Theorem \ref{t:main}) of this paper. 

\bibliographystyle{plain}
\bibliography{../../../../references/references,../../../../references/references2}


\end{document}